\documentclass[prl,aps,twocolumn,superscriptaddress]{revtex4-1}

\usepackage{bm,bbm,color,float,dcolumn,amsmath,amssymb,graphicx,subfigure}

\usepackage[colorlinks=true]{hyperref}
\hypersetup{linkcolor=blue,citecolor=blue,urlcolor=blue}

\graphicspath{ {./figures/} }

\definecolor{psychedelicpurple}{rgb}{0.87, 0.0, 1.0}
\definecolor{violet}{rgb}{0.56, 0, 1}

\begin{document}

\title{Universal scaling of Klein bottle entropy near conformal critical points}
\author{Yueshui Zhang}
\affiliation{Institute of Physics, Chinese Academy of Sciences, Beijing 100190, China}
\affiliation{University of Chinese Academy of Sciences, Beijing 100049, China}

\author{Anton Hulsch}
\affiliation{Institut f\"ur Theoretische Physik, Technische Universit\"at Dresden, 01062 Dresden, Germany}

\author{Hua-Chen Zhang}
\affiliation{Institut f\"ur Theoretische Physik, Technische Universit\"at Dresden, 01062 Dresden, Germany}
\affiliation{Department of Physics and Astronomy, Aarhus University, DK-8000 Aarhus C, Denmark}

\author{Wei Tang}
\affiliation{Department of Physics and Astronomy, Ghent University, Krijgslaan 281, S9, B-9000 Ghent, Belgium}

\author{Lei Wang}
\affiliation{Institute of Physics, Chinese Academy of Sciences, Beijing 100190, China}
\affiliation{Songshan Lake Materials Laboratory, Dongguan, Guangdong 523808, China}

\author{Hong-Hao Tu}
\email{hong-hao.tu@tu-dresden.de}
\affiliation{Institut f\"ur Theoretische Physik, Technische Universit\"at Dresden, 01062 Dresden, Germany}

\date{\today}

\begin{abstract}
We show that the Klein bottle entropy [\href{https://journals.aps.org/prl/abstract/10.1103/PhysRevLett.119.261603}{Phys. Rev. Lett. 119, 261603 (2017)}] for conformal field theories (CFTs) perturbed by a relevant operator is a universal function of the dimensionless coupling constant. The universal scaling of the Klein bottle entropy near criticality provides an efficient approach to extract the scaling dimension of lattice operators via data collapse. As paradigmatic examples, we validate the universal scaling of the Klein bottle entropy for Ising and $\mathbb{Z}_3$ parafermion CFTs with various perturbations using numerical simulation with continuous matrix product operator approach.
\end{abstract}

\maketitle

{\em Introduction} --- The study of continuous phase transitions and critical phenomena is an evergreen topic in theoretical physics~\cite{Cardy1996,Sachdev2011,Vojta2003}. As the correlation length is diverging, systems at and near criticality are described by few variables (e.g., order parameters) that vary slowly in space and time. Their physical properties are insensitive to microscopic details and exhibit universal behaviors shared by different models. From a theoretical perspective, these systems, in the low-energy, long-wavelength limit, are amenable to field theory descriptions.

Given a microscopic model at or near criticality, an important task is to pinpoint the underlying field theory description. While this task can be tremendously difficult, various methods have been made available for one-dimensional (1D) quantum systems and, equivalently, two-dimensional (2D) classical statistical models. The scaling limit of such systems is often described by 2D conformal field theories (CFTs)~\cite{Belavin1984,Francesco1997,Henkel1999}, possibly with additional perturbations that are relevant, marginal, or irrelevant in the renormalization group (RG) sense~\cite{Cardy1986a,Saleur1987,Zamolodchikov1989,Smirnov2017}. As a characteristic quantity of 2D CFTs, the central charge can be read out from the finite-size scaling of the Casimir energy~\cite{Affleck1986,Blote1986} or entanglement entropy in the ground state~\cite{Holzhey1994,Vidal2003,Calabrese2004}.

In many circumstances, the central charge is not enough for distinguishing different CFTs and one calls for a finer distinction. Recently, it is found that 2D CFTs defined on the Klein bottle exhibit a universal entropy~\cite{Tu2017}. This so-called Klein bottle entropy only depends on conformal data (e.g., modular $S$ matrix for rational CFTs~\cite{Tu2017,Tang2017,Chen2017,Wang2018,Garcia2021,Vanhove2022} and the compactification radius for compactified boson CFTs~\cite{Tang2019}) and can be efficiently computed with various numerical methods, making it a competitive tool for characterizing 2D CFTs in numerics (see Refs.~\cite{Li2020,Chen2020,Li2021}).

In this Letter, we extend the scope of Klein bottle entropy from CFT to the scaling region near criticality. By considering a \emph{unitary} CFT perturbed by a \emph{relevant} operator, we show that the Klein bottle entropy, denoted by $K(s)$, is a universal function of some dimensionless coupling $s$. This universal function allows us to extract the conformal weight of the perturbation operator via data collapse. By combining analytical and numerical approaches, we calculate the Klein bottle entropy and verify the universal scaling for Ising and $\mathbb{Z}_3$ parafermion CFTs with various perturbations. Our results clearly suggest that the Klein bottle entropy not only locates critical points accurately, but also provides an efficient method to compute scaling dimension of lattice operators. For latter purpose, current standard method relies on extracting exponents from large-distance correlation functions \emph{at} criticality. In contrast, our present approach deals with \emph{off-critical} systems and uses data collapse, which is numerically less challenging and can provide more accurate estimates.

{\em Universal scaling function} --- We consider the following 1D Hamiltonian describing a perturbed CFT on a circle of length $L$:
\begin{equation}
H = H_{\mathrm{CFT}} - vg \int_0^L \mathrm{d}x \, \varphi(x) \, ,
\label{eq:perturbed-Ham}
\end{equation}
where $H_{\mathrm{CFT}}=\frac{2\pi v}{L}(L_0+\bar{L}_0-\frac{c}{12})$ is the CFT Hamiltonian. Here $L_0$ ($\bar{L}_0$) is the zeroth-level holomorphic (antiholomorphic) Virasoro generator, and $c$ and $v$ are the central charge and velocity, respectively. $g$ is the coupling constant of the perturbation. As the velocity $v$ in Eq.~\eqref{eq:perturbed-Ham} is an overall unit, we set $v = 1$ in field theory analysis and restore it later when analyzing lattice models. We also set $k_{\mathrm{B}} = \hbar = 1$ ($k_{\mathrm{B}}$: Boltzmann's constant) throughout this work. The operator $\varphi$ has conformal weight $(h,\bar{h})$ and, for simplicity, we assume it has a vanishing conformal spin (i.e., $h=\bar{h}$). The normalization of the operator $\varphi$ is fixed by $\lim_{x\rightarrow \infty}\lim_{L\rightarrow \infty} x^{4h} \langle \varphi(0) \varphi(x) \rangle = 1$, where the expectation value is taken in the vacuum of the CFT. For $h<1$, the perturbation is RG relevant and drives the system away from criticality.

At inverse temperature $\beta$, the partition function $Z^{\mathcal{T}}(L,\beta,g) = \mathrm{tr}(e^{-\beta H})$ lives on a torus, and the trace can be evaluated with the eigenstates of $H_{\mathrm{CFT}}$, denoted by $|\alpha, \bar{\gamma}\rangle$. The Klein bottle partition function $Z^{\mathcal{K}}(L,\beta,g) = \mathrm{tr}(\Omega e^{-\beta H})$ has an extra operator $\Omega$ satisfying $\Omega^2 = 1$. In the path integral picture, $\Omega$ has the intuitive meaning of gluing the ``field configurations'' (before and after the imaginary-time evolution) in a spatially inverted fashion and hence changes the manifold from torus to Klein bottle. Rigorously speaking, there are various choices of $\Omega$ for a self-consistent definition of the Klein bottle partition function~\cite{Felder2002}, and we choose the one whose action on the eigenstates of $H_{\mathrm{CFT}}$ is $\Omega |\alpha, \bar{\gamma}\rangle = |\gamma, \bar{\alpha}\rangle$. For many lattice models, one can realize this by simply choosing spatial reflection~\cite{Tu2017,Tang2017,Chen2017}.

We are interested in the Klein bottle entropy, defined as the ratio of two partition functions
\begin{equation}
K(\beta,g) = \lim_{L \rightarrow \infty} \frac{Z^{\mathcal{K}}(2L,\frac{\beta}{2},g)}{Z^{\mathcal{T}}(L,\beta,g)} \, .
\label{eq:KB-entropy}
\end{equation}
At the critical point ($g = 0$), $K$ acquires a universal value (i.e., independent of $\beta$ and velocity $v$) and can be used for distinguishing different CFTs~\cite{Tu2017,Tang2017,Chen2017,Tang2019,Vanhove2022}. When moving away from criticality, it is natural to ask how the Klein bottle entropy $K(\beta,g)$ varies as a function of $g$ and $\beta$.

In the limit $L \rightarrow \infty$, $\beta$ is the only length scale in the theory. Since the Hamiltonian $H$ has dimension $[\beta]^{-1}$ and the operator $\varphi$ has dimension $[\beta]^{-2h}$, dimension of the coupling constant $g$ is $[\beta]^{2h-2}$, and a dimensionless coupling $s=g \beta^{2-2h}$ is hence the only parameter in the theory, measuring the strength of the perturbation. Since the Klein bottle entropy \eqref{eq:KB-entropy} is also dimensionless, it must be a \emph{universal} function of the dimensionless coupling $s$, denoted as $K(s)$.

The Klein bottle entropy $K(s)$ being universal both at and near criticality has immediate applications in numerical studies: (i) It establishes a firm foundation for using the Klein bottle entropy of the CFT to locate conformal critical points. (ii) Using lattice operators as ``probe perturbations,'' one can exploit data collapse of the Klein bottle entropy to accurately determine conformal weights of lattice operators.

\begin{figure}
\resizebox{\columnwidth}{!}{\includegraphics{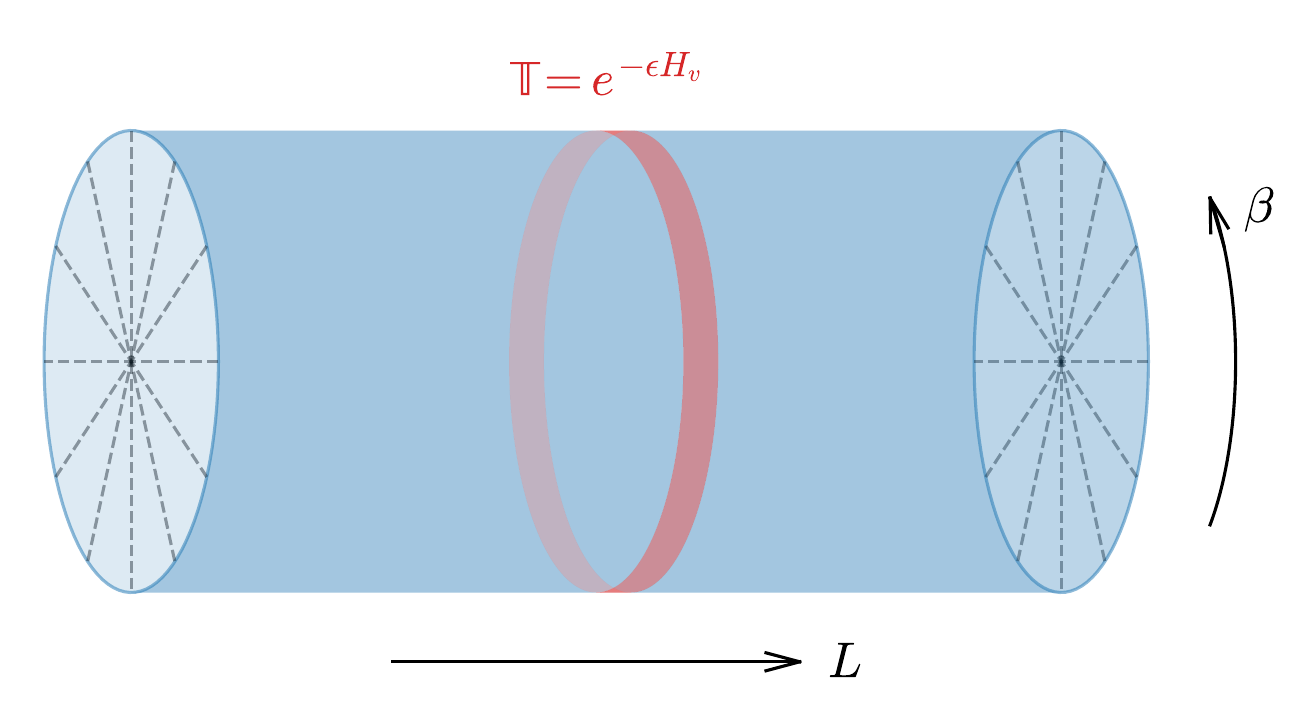}}
\caption{Klein bottle partition function $Z^{\mathcal{K}}(2L,\frac{\beta}{2},g)$ formulated on a cylinder with length $L$, circumference $\beta$, and crosscap boundaries. The (light red) thin ribbon represents the quantum transfer matrix $\mathbbm{T}(\beta,g)$ generating evolution along the spatial direction. }
\label{fig:crosscap}
\end{figure}

It is worth mentioning that the quantum transfer matrix gives an alternative perspective of the universal Klein bottle entropy. By using a ``cut-and-sew'' procedure~\cite{Tang2017}, the Klein bottle with size $(2L,\beta/2)$ is mapped to a cylinder with length $L$, circumference $\beta$, and ``crosscap'' boundaries (see Fig.~\ref{fig:crosscap}). In this picture, the Klein bottle partition function
\begin{equation}
Z^{\mathcal{K}}(2L,\frac{\beta}{2},g) = \langle \mathcal{C}|e^{-LH_v(\beta,g)}|\mathcal{C}\rangle
\label{eq:crosscap-Z}
\end{equation}
is viewed as a spatial evolution generated by the quantum transfer matrix $\mathbbm{T}(\beta,g) \equiv e^{-\epsilon H_v(\beta,g)}$ ($\epsilon$: short-distance cutoff) between two crosscap boundary states, $|\mathcal{C}\rangle$ and its conjugate. For space-time symmetric theories addressed in this work, $H_v$ takes the same form as (\ref{eq:perturbed-Ham}), except that the imaginary time $\tau$ (inverse temperature $\beta$) plays the role of the spatial coordinate $x$ (length $L$). Using $e^{-\epsilon E(\beta,g)}$ ($|\psi(\beta,g)\rangle$) to denote the leading eigenvalue (normalized leading eigenvector) of $\mathbbm{T}(\beta,g)$, the evolution of the quantum transfer matrix in \eqref{eq:crosscap-Z}, for $L \gg \beta$, projects onto the leading eigenvector of $\mathbbm{T}$:
\begin{equation}
Z^{\mathcal{K}}(2L,\frac{\beta}{2},g) \simeq e^{-E(\beta,g) L}|\langle \mathcal{C}|\psi(\beta,g) \rangle|^2 \, .
\end{equation}
We note that $|\psi(\beta,g) \rangle$ as the ground state of $H_v(\beta,g)$, which describes the same theory as \eqref{eq:perturbed-Ham} but is defined on a circle of length $\beta$, only depends on the dimensionless coupling $s=g \beta^{2-2h}$ and can hence be written as $|\psi(\beta,g) \rangle \equiv |\psi(s) \rangle$. Similarly, the torus partition function is evaluated as $Z^{\mathcal{T}}(L,\beta,g) = \mathrm{tr}[e^{-LH_v(\beta,g)}] \simeq e^{-E(\beta,g) L}$ for $L \gg \beta$. Using these results, the Klein bottle entropy \eqref{eq:KB-entropy} is simplified as
\begin{equation}
K(s) = |\langle \mathcal{C}|\psi(s) \rangle|^2 \, ,
\label{eq:crosscap-entropy}
\end{equation}
which reaffirms $K(s)$ is universal. For 1D quantum Hamiltonians, the continuous matrix product operator (cMPO) method~\cite{Tang2020} provides an efficient way to compute the Klein bottle entropy via Eq.~\eqref{eq:crosscap-entropy}.

Although the Klein bottle entropy being universal has very promising prospects, the calculation of its full analytical form is quite challenging. Nevertheless, there are at least two routes to proceed: (i) Treat the $\varphi$ term in \eqref{eq:perturbed-Ham} perturbatively and develop the perturbation theory based on \eqref{eq:crosscap-Z}. This effectively generates a series expansion $K(s) = \sum_{n=0}^{\infty} K_n s^n$, where $K_0$ is just the Klein bottle entropy of the CFT, and higher-order coefficients $K_{n>0}$ are obtained order by order in the perturbation theory. However, one has to be cautious about the possible nonanalyticity of $K(s)$ at $s=0$, which might cause divergences in the perturbative expansion. (ii) For integrable field theories, the overlap between Bethe vectors and the crosscap boundary state, which gives the Klein bottle entropy via Eq.~\eqref{eq:crosscap-entropy}, might be calculable~\cite{Caetano2022,Ekman2022,Gombor2022a,Gombor2022b}.

{\em Perturbed Ising CFT} --- As a concrete example, we consider the perturbed Ising CFT:
\begin{equation}
H = H_{\mathrm{Ising}} - g_1 \int_0^L\mathrm{d}x \, \varepsilon(x) - g_2 \int_0^L\mathrm{d}x \, \sigma(x) \, ,
\label{eq:perturbed-Ising}
\end{equation}
where $H_{\mathrm{Ising}}$ is the Hamiltonian of the Ising CFT with central charge $c=1/2$ and velocity $v=1$. $\varepsilon$ and $\sigma$ are primary fields of the Ising CFT with conformal weight $(1/2,1/2)$ and $(1/16,1/16)$, respectively. This field theory is known to describe the scaling limit of the 2D classical Ising model~\cite{Zamolodchikov1989,Mussardo2020}, where $\varepsilon$ ($\sigma$) corresponds to thermal (magnetic) perturbation. For $g_2 = 0$, the field theory \eqref{eq:perturbed-Ising} can also be formulated with a free Majorana fermion, where the $\varepsilon$ term becomes the mass of the Majorana fermion. This allows us to derive an exact expression for the Klein bottle entropy~\cite{Appendix}
\begin{equation}
K(s_1) = 1 + \frac{1}{\sqrt{1+e^{2\pi s_1}}}
\label{eq:Ising-Ks1}
\end{equation}
with dimensionless coupling $s_1 = g_1 \beta$. At the critical point ($s_1 = 0$), $K(0) = 1 + \frac{\sqrt{2}}{2}$ restores the Klein bottle entropy of the Ising CFT~\cite{Tu2017}. In two limits ($s_1 \rightarrow -\infty$ and $+\infty$), $K(-\infty) = 2$ ($K(+\infty) = 1$) reflects the twofold degenerate (unique) ground state in the Ising ordered (disordered) phase. For $g_1 = 0$, we are unable to derive an analytical expression for the Klein bottle entropy $K(s_2)$ (dimensionless coupling: $s_2 = g_2 \beta^{15/8}$) and have to resort to numerical approaches.

For numerical simulations, we consider the following 1D quantum Ising chain with both transverse and longitudinal fields:
\begin{equation}
H = -\sum_{j=1}^N \sigma^x_j\sigma^x_{j+1} - h_1 \sum_{j=1}^N\sigma^z_j - h_2 \sum_{j=1}^N\sigma^x_j \, ,
\label{eq:quantum-Ising}
\end{equation}
where $\sigma^{\alpha}_j$ ($\alpha=x,z$) are Pauli spin operators at site $j$, $N$ is the total number of sites, and periodic boundary condition ($\sigma^{\alpha}_{N+j} = \sigma^{\alpha}_j$) is imposed. The Ising CFT is realized at $h_1 = 1$ and $h_2 = 0$ with velocity $v=2$ (lattice spacing set to unity here and hereafter). The Klein bottle partition function on the lattice is defined by $Z^{\mathcal{K}} = \mathrm{tr}(P e^{-\beta H})$, where $P$ is the spatial reflection operator whose action on the Ising spin basis is given by $P|\sigma_1,\sigma_2,\ldots,\sigma_{N} \rangle = |\sigma_N,\ldots,\sigma_{2},\sigma_{1} \rangle$ with $\sigma_j = \pm 1$.

To compare lattice and field theory results, one should take into account the velocity as well as the normalization of perturbation operators on the lattice. Similar to the field theory prescription, the normalization of operators is obtained from two-point correlators in the critical ground state: $\mathcal{N}_{\varepsilon}=\lim_{r\rightarrow \infty}\lim_{N\rightarrow \infty} r^{2} \langle \sigma^z_j \sigma^z_{j+r} \rangle_{\mathrm{c}} =1/\pi^2$ and $\mathcal{N}_\sigma=\lim_{r\rightarrow \infty}\lim_{N\rightarrow \infty} r^{1/4} \langle \sigma^x_j \sigma^x_{j+r} \rangle_{\mathrm{c}} \approx 0.645$~\cite{Pfeuty1970}, where $\langle \cdots \rangle_{\mathrm{c}}$ denotes connected correlators with local expectation values subtracted. Taking into account the velocity, dimensionless couplings for the lattice model \eqref{eq:quantum-Ising} are given by $s_1 = \frac{\sqrt{\mathcal{N}_{\varepsilon}}}{v}(h_1-1) v\beta$ and $s_2 =\frac{\sqrt{\mathcal{N}_\sigma}}{v}h_2(v\beta)^{15/8}$. However, if one simply aims at determining conformal weights from the universal scaling (rather than quantitative comparison with field theory calculations), it suffices to use, e.g., $\tilde{s}_1 = (h_1-1)\beta$ and $\tilde{s}_2 = h_2 \beta^{15/8}$ without the extra rescaling.

\begin{figure}
\resizebox{\columnwidth}{!}{\includegraphics{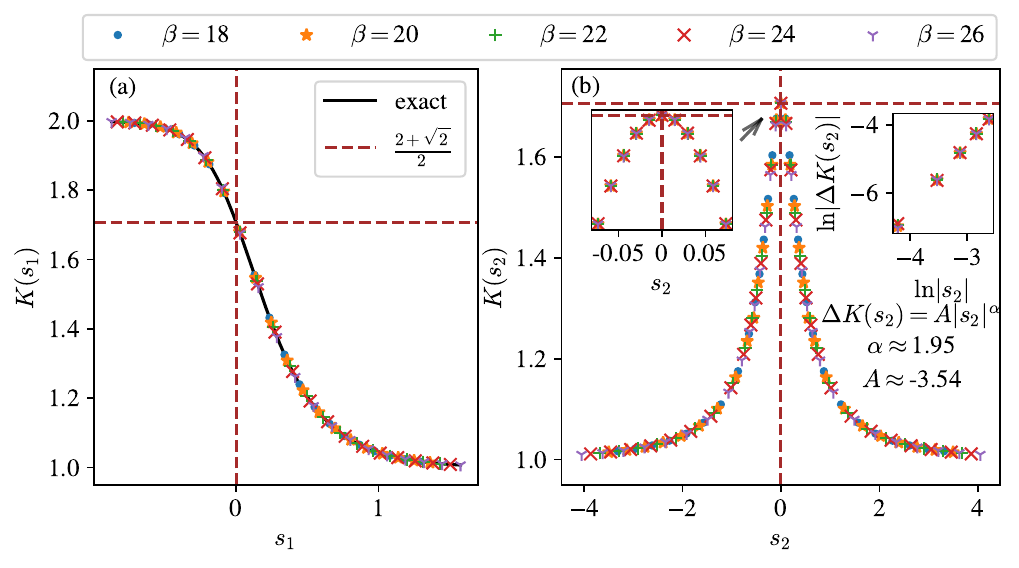}}
\caption{Klein bottle entropy for the Ising CFT with (a) thermal and (b) magnetic perturbations. The numerical data are obtained from the cMPO calculation (bond dimension $\chi=20$) with the quantum Ising chain \eqref{eq:quantum-Ising}. The field theory result for $K(s_1)$ [Eq.~\eqref{eq:Ising-Ks1}] is shown as the solid line in (a). Horizontal and vertical dashed lines indicate the critical point and the Klein bottle entropy of the Ising CFT, respectively. Left inset of (b): Enlargement near $s_2=0$. Right inset of (b): Fitting the data near $s_2=0$ with $\Delta K(s_2) = K(s_2) - K(0) \approx A|s_2|^{\alpha}$.
}
\label{fig:ising}
\end{figure}

We calculate the Klein bottle entropy numerically for the quantum Ising chain \eqref{eq:quantum-Ising} using the cMPO method~\cite{Tang2020} and plot the results in Fig.~\ref{fig:ising}. For the case of thermal perturbation, numerical data shown in Fig.~\ref{fig:ising}(a) agree very well with the analytical result in Eq.~\eqref{eq:Ising-Ks1}, which confirms the universality of $K(s_1)$. For the case of magnetic perturbation, the data collapse is also excellent [see Fig.~\ref{fig:ising}(b)] and the fitting $K(s_2) = K(0) + A |s_2|^{\alpha}$ near $s_2=0$ yields $\alpha \approx 1.95$ and $A \approx -3.54$ [inset of Fig.~\ref{fig:ising}(b)]. The exact value for $\alpha$ is expected to be $2$, which is the second-order term in the series expansion of $K(s_2)$. The field theory calculation of the (universal) coefficient $A$ is an interesting task for a future work. When both thermal and magnetic perturbations are present, we have numerically calculated $K(s_1,s_2)$ and also observed excellent data collapse (not shown).

{\em Perturbed $\mathbb{Z}_3$ parafermion CFT} --- As the second example, we consider the perturbed $\mathbb{Z}_3$ parafermion CFT:
\begin{equation}
H = H_{\mathrm{Parafermion}} - g \int_0^L\mathrm{d}x \, \varepsilon(x) \, ,
\label{eq:perturbed-parafermion}
\end{equation}
where $H_{\mathrm{Parafermion}}$ is the Hamiltonian of the $\mathbb{Z}_3$ parafermion CFT with central charge $c=4/5$. The torus partition function of the $\mathbb{Z}_3$ parafermion CFT is the nondiagonal modular invariant of the $\mathcal{M}(6,5)$ minimal model~\cite{Francesco1997}. The field theory \eqref{eq:perturbed-parafermion} describes the scaling limit of the 2D classical three-state Potts model, where the operator $\varepsilon$ is a primary field of the $\mathcal{M}(6,5)$ minimal model with conformal weight $(2/5,2/5)$ and corresponds to the thermal perturbation. Different from the Ising case, $\mathbb{Z}_3$ parafermion CFT is an interacting theory without free-field representation. In the presence of thermal perturbation, the Klein bottle entropy is difficult to calculate directly from the field theory.

Here we consider the lattice realization of \eqref{eq:perturbed-parafermion} in the three-state quantum clock chain
\begin{equation}
H = -\sum_{j=1}^N (\sigma^{\dag}_j\sigma_{j+1} + \sigma^{\dag}_{j+1}\sigma_{j}) - h_3 \sum_{j=1}^N (\tau_j + \tau^{\dag}_j)  \, ,
\label{eq:quantum-clock}
\end{equation}
where
\begin{align}
\sigma = \begin{pmatrix}
0 & 1 & 0  \\
0 & 0 & 1 \\
1 & 0 & 0
\end{pmatrix},  \quad
\tau = \begin{pmatrix}
1 & 0 & 0  \\
0 & e^{2\pi \mathrm{i}/3} & 0 \\
0 & 0 & e^{4\pi \mathrm{i}/3}
\end{pmatrix},
\end{align}
are $\mathbb{Z}_3$ spin matrices. The $\mathbb{Z}_3$ parafermion CFT describes the critical point of \eqref{eq:quantum-clock} at $h_3 = 1$, and the velocity is $v=\frac{3\sqrt{3}}{2}$~\cite{Albertini1992}. The normalization of the perturbation operator is obtained by numerically calculating the correlator at the critical point using the variational uniform matrix product state method~\cite{Zauner2018,mpskit}: $\mathcal{N}_{\varepsilon}=\lim_{r\rightarrow \infty}\lim_{N\rightarrow \infty} r^{8/5} \langle (\tau^{\dag}_j + \tau_j) (\tau^{\dag}_{j+r} + \tau_{j+r}) \rangle_{\mathrm{c}} \approx 0.315$. Taking into account the velocity and the normalization of the perturbation operator, the dimensionless coupling $s$ for the lattice model \eqref{eq:quantum-clock} is defined as $s = \frac{\sqrt{\mathcal{N}_{\varepsilon}}}{v} (h_3-1) (v\beta)^{6/5}$.

\begin{figure}
\centering
\resizebox{0.75\columnwidth}{!}{\includegraphics{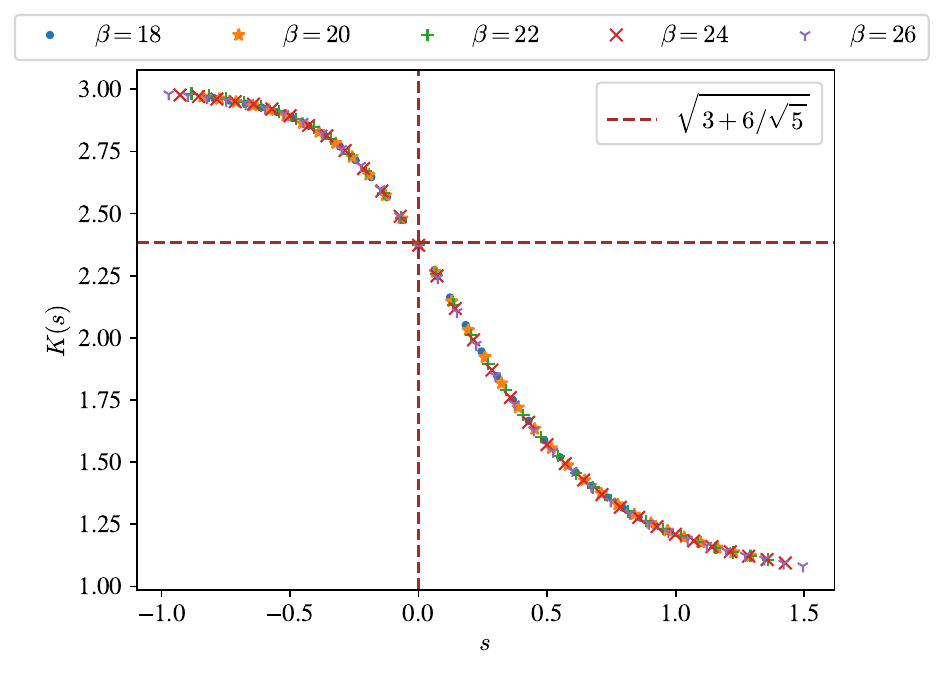}}
\caption{The Klein bottle entropy $K(s)$ as a function of the dimensionless coupling $s$ for the three-state quantum clock chain near criticality. The numerical data are obtained with the cMPO approach (bond dimension $\chi=24$) for different $\beta$ (shown in legend). The critical point and the Klein bottle entropy of the $\mathbb{Z}_3$ parafermion CFT are indicated by horizontal and vertical dashed lines, respectively.}
\label{fig:potts}
\end{figure}

The Klein bottle entropy for the three-state quantum clock chain \eqref{eq:quantum-clock} has been calculated using the cMPO method and the results are shown in Fig.~\ref{fig:potts}. The data collapse for different $\beta$ is again observed. The numerical result at the critical point ($s=0$) agrees very well with the expected Klein bottle entropy of the $\mathbb{Z}_3$ parafermion CFT, $K(0) = \sqrt{3 + 6/\sqrt{5}}$~\cite{Tang2017}. In two limits ($s \rightarrow -\infty$ and $+\infty$), $K(-\infty) = 3$ ($K(+\infty) = 1$) indicates the three-fold degenerate (unique) ground state in the $\mathbb{Z}_3$ symmetry breaking (disordered) phase.

The analytical and numerical results for perturbed Ising and $\mathbb{Z}_3$ parafermion CFTs give some hints on possible general features of the Klein bottle entropy. The CFT under a relevant perturbation (with dimensionless coupling $s$) separates two off-critical phases, denoted by $A$ ($s<0$) and $B$ ($s>0$). There are two typical scenarios: (i) Phase $A$ ($B$) has a broken (an unbroken) discrete symmetry with (without) ground-state degeneracy. The Klein bottle entropy $K(-\infty)$ (an integer greater than one) equals the number of degenerate ground states in phase $A$, while $K(+\infty) = 1$ signals a unique ground state in phase $B$. In this case, we conjecture that $K(s)$ \emph{monotonically} decreases from $K(-\infty)$ to 1 as $s$ increases from $-\infty$ to $+\infty$. (ii) If neither phase $A$ nor $B$ spontaneously breaks a symmetry, the ground state is unique in both phases, indicated by the Klein bottle entropy $K(\pm \infty) = 1$. In this situation, we conjecture that $K(s)$ achieves its maximum at the critical point $s=0$ and \emph{monotonically} decreases as $s$ increases (decreases) from zero to $+\infty$ ($-\infty$). In addition to the present work (as well as numerical evidences in Refs.~\cite{Chen2017,Li2020}), rigorous results~\cite{Caetano2022} obtained from certain integrable field theories (staircase model~\cite{Zamolodchikov2006} and its generalization) also support these conjectures.

{\em Discussion} --- In summary, we have shown that for conformal critical points perturbed by a relevant operator, the Klein bottle entropy $K(s)$ is a universal function of the dimensionless coupling constant $s$. This allows us to devise an efficient method to determine the conformal weight of perturbation operators via data collapse. The analytic and numerical results of Ising and $\mathbb{Z}_3$ parafermion CFTs with various perturbations illustrate an excellent agreement with the prediction.

The universal scaling of the Klein bottle entropy has opened a new venue in the study of 2D field theories. To proceed, a plausible direction is to develop methods for computing the universal scaling function $K(s)$. For instance, it should be possible to establish a conformal perturbation theory to calculate the leading-order terms in the series expansion of $K(s)$. The exact form of $K(s)$ might also be extracted for some integrable field theories or spin chains~\cite{Caetano2022,Ekman2022,Gombor2022a,Gombor2022b}.

For future works, it would be interesting to study, both perturbatively and nonperturbatively, under which conditions $K(s)$ decreases or increases monotonically. The answer to this question might uncover a deep relation between the Klein bottle entropy and the bulk RG flow, analogous to Zamolodchikov's $c$ theorem~\cite{Zamolodchikov1986}. Apart from the relevant perturbations considered in this work, it is desirable to study the effect of marginal perturbations on the Klein bottle entropy, too. Needless to say, it would be fruitful if a suitable generalization of the Klein bottle entropy could be found in higher dimensions. As a higher-dimensional CFT perturbed by a relevant operator is also controlled by the dimensionless coupling, a \emph{dimensionless} entropy, possibly arising on certain closed manifold, would be a universal function of the dimensionless coupling and can characterize the critical theory. Considering the wide adoption of an alternative dimensionless ratio, the Binder cumulant of order parameters, in the study of critical phenomena~\cite{Binder1981}, we believe further exploration of Klein bottle entropy will be fruitful.

{\em Acknowledgments} --- We are grateful to Meng Cheng, J\"urgen Fuchs, Masaki Oshikawa, and Christoph Schweigert for helpful discussions. H.-H.T. would like to thank the Erwin Schr\"odinger International Institute for Mathematics and Physics at the University of Vienna for support and hospitality during the thematic programme \textit{Tensor Networks: Mathematical Structures and Novel Algorithms}, where part of this work has been carried out. This work is supported by the Strategic Priority Research Program of the Chinese Academy of Sciences under Grant No.~XDB30000000 and National Natural Science Foundation of China under Grants No.~92270107, No.~T2225018 and No.~T2121001, the European Research Council (ERC) under the European Union's Horizon 2020 research and innovation programme (Grant Agreements No.~715861 (ERQUAF)), and the Deutsche Forschungsgemeinschaft (DFG) through project A06 of SFB 1143 (Project No.~247310070).

\bibliography{refs}

\clearpage
\onecolumngrid

\section*{Supplemental Material}

\setcounter{table}{0}
\renewcommand{\thetable}{S\arabic{table}}
\setcounter{figure}{0}
\renewcommand{\thefigure}{S\arabic{figure}}
\setcounter{equation}{0}
\renewcommand{\theequation}{S\arabic{equation}}

\appendix

\section{Ising CFT with thermal perturbation}

\subsection{Majorana fermion field theory}

The Ising CFT with thermal perturbation can be represented as a free Majorana fermion field theory with mass. In the Hamiltonian formulation, the system is defined on a circle with length $L$:
\begin{align}
H &= H_{\mathrm{Ising}} - g_1 \int_0^L\mathrm{d}x \, \varepsilon(x) \nonumber \\
&= \frac{i}{2} \int_0^L \mathrm{d}x \, [\chi(x) \partial_x \chi(x) - \bar{\chi}(x) \partial_x \bar{\chi}(x)] - im \int_0^L \mathrm{d}x \, \chi(x) \bar{\chi}(x) \, ,
\label{eq:Major-QFT}
\end{align}
where $\chi$ ($\bar{\chi}$) describes a left-moving (right-moving) Majorana fermion. The Majorana operators are self-conjugate, $\chi^\dagger=\chi$ (similar for $\bar{\chi}$), and they satisfy the anticommutation relations, $\{\chi(x),\chi(x')\} = \{\bar{\chi}(x),\bar{\chi}(x')\} =\delta(x-x')$ and $\{\chi(x),\bar{\chi}(x')\} = 0$. The velocity in \eqref{eq:Major-QFT} has already been set to unity. The energy operator $\varepsilon$ has conformal weight $h=\bar{h}=\frac{1}{2}$ and is represented with Majorana operators as $\varepsilon(x) = 2\pi i \chi(x)\bar{\chi}(x)$, where the prefactor $2\pi$ ensures that the normalization agrees with the adopted convention in the main text. Thus, the dimensionless coupling in \eqref{eq:Major-QFT} is given by
\begin{equation}
s = g_1\beta^{2-2h} = \frac{m\beta}{2\pi} \, .
\end{equation}
Note that this is defined as $s_1$ in the main text, and we drop the subscript here since no confusion would arise.

The Fourier transform of $\chi(x)$ and its inverse are given by
\begin{equation}
\chi(x) = \frac{1}{\sqrt{L}} \sum_k \chi_k e^{ikx}, \qquad \chi_k = \frac{1}{\sqrt{L}} \int^L_0 \mathrm{d}x \, \chi(x) e^{-ikx} \, .
\label{eq:Fourier}
\end{equation}
The definition for $\bar{\chi}$ is similar, which we would omit for now. Both antiperiodic and periodic boundary conditions of Majorana operators, defined as $\chi(x+L) = -\chi(x)$ and $\chi(x+L) = \chi(x)$, should be considered in Ising field theory. These are known as the Neveu-Schwarz (NS) and Ramond (R) sectors, respectively. In the NS (R) sector, the allowed momenta in Eq.~\eqref{eq:Fourier} are $k = \frac{2\pi}{L}(n - \frac{1}{2})$ ($k = \frac{2\pi}{L}n$) with $n \in \mathbb{Z}$. The Majorana operator $\chi(x)$ being self-conjugate implies $\chi_{-k} = \chi_k^{\dag}$. We note that $\chi_{k=0}$ in the R sector is of Majorana nature (up to normalization), namely, $\chi_0^{\dag} = \chi_0$ and $\chi_0^2 = 1/2$.

By using the Fourier transform of $\chi(x)$ and $\bar{\chi}(x)$, the Hamiltonian \eqref{eq:Major-QFT} is diagonalized. In the R sector, we have
\begin{align}
H_{\mathrm{R}} &= \frac{1}{2}\sum_k k(-\chi_{-k}\chi_k + \bar{\chi}_{-k}\bar{\chi}_k) - im\sum_{k}\chi_{-k}\bar{\chi}_k \nonumber \\
& = \sum_{k>0} k(-\chi_{-k}\chi_k+\bar{\chi}_{-k}\bar{\chi}_k)-im\sum_{k>0}(\chi_{-k}\bar{\chi}_k-\bar{\chi}_{-k}\chi_k) -im\chi_0\bar{\chi}_0 \nonumber \\
&=\sum_{k>0}\begin{pmatrix}\chi_{-k} &\bar{\chi}_{-k}\end{pmatrix}
\begin{pmatrix}
-k & -im\\im & k
\end{pmatrix}
\begin{pmatrix}
\chi_k\\ \bar{\chi}_k
\end{pmatrix}-im\chi_0\bar{\chi}_0 \nonumber \\
&=\sum_{k> 0}\sqrt{k^2+m^2}(\eta_k^\dagger\eta_k-\eta_{-k}\eta_{-k}^\dagger)-im\frac{1}{\sqrt{2}}(\eta_0+\eta_0^\dagger) \cdot \frac{i}{\sqrt{2}}(\eta_0-\eta_0^\dagger) \nonumber \\
&=\sum_{k\neq 0}\sqrt{k^2+m^2}(\eta_k^\dagger\eta_k-\frac{1}{2})+m(\eta_0^\dagger\eta_0-\frac{1}{2}) \nonumber \\
&=\sum_{k\in\mathrm{R}}\epsilon_k(\eta_k^\dagger\eta_k-\frac{1}{2})
\label{eq:H_R}
\end{align}
with single-particle energy $\epsilon_k=\sqrt{k^2+m^2}$ for $k\neq 0$ and $\epsilon_0=m$. For $k=0$, we have defined a (complex) fermionic mode $\eta_0=\frac{1}{\sqrt{2}}(\chi_0-i\bar{\chi}_0)$. The diagonalization in the NS sector is similar and yields $H_{\mathrm{NS}}  =\sum_{k\in\mathrm{NS}}\epsilon_k(\eta_k^\dagger\eta_k-\frac{1}{2})$.

The vacuum in the NS (R) sector, written as $|0\rangle_{\mathrm{NS}}$ ($|0\rangle_{\mathrm{R}}$), is annihilated by all $\eta_{k \in \mathrm{NS}(\mathrm{R})}$ modes. The vacuum energies are given by
\begin{align}
E_0^{\mathrm{NS}} = -\frac{1}{2}\sum_{k\in\mathrm{NS}} \epsilon_k
= -\sum_{n=1}^\infty \frac{2\pi}{L}\sqrt{(n-\frac{1}{2})^2+t^2} \, , \label{eq:E_NS}
\\
E_0^{\mathrm{R}} = -\frac{1}{2}\sum_{k\in\mathrm{R}} \epsilon_k
= -\frac{m}{2}-\sum_{n=1}^\infty \frac{2\pi}{L}\sqrt{n^2+t^2} \, ,
\label{eq:E_R}
\end{align}
where $t=\frac{mL}{2\pi}$ is the dimensionless mass.

The vacuum energies can be calculated by expanding the mass (i.e., treating the mass perturbatively). In the NS sector, we obtain
\begin{align}
E_0^{\mathrm{NS}} &=-\frac{2\pi}{L}\sum_{n=1}^\infty (n-\frac{1}{2})\cdot \sqrt{1+\left( \frac{t
}{n-\frac{1}{2}}\right) ^{2}} \nonumber \\
&=-\frac{2\pi}{L}\sum_{n=1}^\infty(n-\frac{1}{2})\cdot \sum_{l=0}^\infty C_{\frac{1}{2}}^l \left(\frac{t}{n-\frac{1}{2}}\right)^{2l} \nonumber \\
&=-\frac{2\pi}{L}\sum_{n=1}^\infty (n-\frac{1}{2})
-t^2 \cdot \frac{\pi}{L} \sum_{n=1}^\infty\frac{1}{n-\frac{1}{2}}
-\frac{2\pi}{L}\sum_{l=2}^\infty t^{2l} C_{\frac{1}{2}}^l \sum_{n=1}^\infty\frac{1}{(n-\frac{1}{2})^{2l-1}} \nonumber \\
&=-\frac{2\pi}{L}\sum_{n=1}^\infty (n-\frac{1}{2})
-t^2 \cdot \frac{\pi}{L}\sum_{n=1}^\infty\frac{1}{n-\frac{1}{2}}
-\frac{2\pi}{L}\sum_{l=2}^\infty t^{2l} C_{\frac{1}{2}}^l (2^{2l-1}-1)\zeta(2l-1) \, ,
\label{eq:NS-vacuum-energy}
\end{align}
where $C^l_{1/2}$ is the binomial coefficient, $C^l_{1/2} \equiv \binom{1/2}{l} = \binom{2l}{l}\frac{(-1)^{l+1}}{2^{2l}(2l-1)}$, and $\zeta$ is the Riemann-Zeta function. The first two terms in Eq.~\eqref{eq:NS-vacuum-energy} are divergent and should be regularized. Below we use the exponential regularization with a short-distance cutoff $a$ (of the order of lattice spacing) to calculate the first two terms:
\begin{align}
\frac{2\pi}{L}\sum_{n=1}^\infty \left(n-\frac{1}{2}\right)
\rightarrow
\sum_{k\in\mathrm{NS},k>0}ke^{-ka}
=-\frac{\partial}{\partial a} \sum_{n=1}^\infty e^{-\frac{2\pi}{L}(n-\frac{1}{2})a}
=\frac{\pi}{2L}\frac{\cosh (\frac{\pi}{L}a)}{\sinh^2 (\frac{\pi}{L}a)}
=\frac{L}{2\pi a^2}+\frac{\pi}{12L}+\mathcal{O}(a^2)
\end{align}
and
\begin{align}
\sum_{n=1}^\infty\frac{1}{n-\frac{1}{2}}
&\rightarrow
\frac{2\pi}{L} \sum_{k\in \mathrm{NS},k>0}\frac{1}{k}e^{-ka}
=\sum_{n=1}^\infty\frac{1}{n-\frac{1}{2}}e^{-\frac{2\pi}{L}(n-\frac{1}{2})a}
= 2\tanh^{-1}(e^{-\frac{\pi}{L}a})
=2\ln 2-\ln\left(\frac{2\pi}{L}a\right)+\mathcal{O}(a^2)  \, .
\end{align}
Using the above regularized results (i.e., dropping the ultraviolet divergent term $L/2\pi a^2$), we obtain
\begin{align}
E_0^{\mathrm{NS}} \rightarrow \frac{2\pi}{L}\left[-\frac{1}{24}+\frac{t^2}{2}\ln\left(\frac{\pi}{2L}a\right) - \sum_{l=2}^\infty t^{2l} C_{\frac{1}{2}}^l (2^{2l-1}-1)\zeta(2l-1)\right]
\equiv \frac{\pi}{L}\gamma_{-}(t)
\end{align}
with
\begin{align}
\gamma_{-}(t) = -\frac{1}{12} + t^2 \ln\left(\frac{\pi}{2L}a\right) - 2\sum_{l=2}^\infty t^{2l} C_{\frac{1}{2}}^l (2^{2l-1}-1)\zeta(2l-1) \, .
\end{align}

In the R sector, we use the same regularization scheme to calculate $E_0^{\mathrm{R}}$ [Eq.~\eqref{eq:E_R}]:
\begin{align}
E_0^{\mathrm{R}}
&=-\frac{m}{2}-\frac{2\pi}{L}\sum_{n=1}^\infty n \cdot \sqrt{1+\left( \frac{t
}{n}\right)^{2}} \nonumber \\
&=-\frac{2\pi}{L}\frac{t}{2}-\frac{2\pi}{L}\sum_{n=1}^\infty n \cdot \sum_{l=0}^\infty C_{\frac{1}{2}}^l \left(\frac{t}{n}\right)^{2l} \nonumber \\
&= -\frac{2\pi}{L}\frac{t}{2} -\frac{2\pi}{L}\sum_{n=1}^\infty n
- t^2 \cdot \frac{\pi}{L}\sum_{n=1}^\infty\frac{1}{n}
-\frac{2\pi}{L}\sum_{l=2}^\infty t^{2l} C_{\frac{1}{2}}^l \sum_{n=1}^\infty\frac{1}{n^{2l-1}} \nonumber \\
&\rightarrow -\frac{2\pi}{L}\frac{t}{2} -\frac{\pi}{2L}\frac{1}{\sinh^2 (\frac{\pi}{L}a)}- t^2 \cdot \frac{\pi}{L} [-\ln(1-e^{-\frac{2\pi}{L}a})]
-\frac{2\pi}{L}\sum_{l=2}^\infty t^{2l} C_{\frac{1}{2}}^l \zeta(2l-1) \nonumber \\
&=-\frac{L}{2\pi a^2}+\frac{2\pi}{L}\left[\frac{1}{12}-\frac{t}{2}+\frac{t^2}{2}\ln\left(\frac{2\pi}{L}a\right)-\sum_{l=2}^\infty t^{2l} C_{\frac{1}{2}}^l \zeta(2l-1) \right] \, .
\end{align}
After dropping the same ultraviolet divergent term $L/2\pi a^2$, we arrive at $ E_0^{\mathrm{R}} \equiv \frac{\pi}{L}\gamma_{+}(t)$ with
\begin{align}
\gamma_{+}(t) = \frac{1}{6} - t + t^2 \ln\left(\frac{2\pi}{L}a\right) - 2\sum_{l=2}^\infty t^{2l} C_{\frac{1}{2}}^l \zeta(2l-1) \, .
\end{align}

The above calculations finish the diagonalization of the Hamiltonian \eqref{eq:Major-QFT}. As a self-consistent check, the result can be verified by calculating the so-called ``universal gap function''~\cite{Henkel1987,Saleur1987,Oshikawa2019}, which is just the finite-size gap between the lowest-energy states in NS and R sectors. However, one has to bear in mind that in the Ising field theory, the NS (R) sector has an even (odd) number of fermions. While $|0\rangle_{\mathrm{NS}}$ is already the lowest-energy state in the NS sector with energy $E_0^{\mathrm{NS}}$, $\eta^{\dag}_{0}|0\rangle_{\mathrm{R}}$ (rather than $|0\rangle_{\mathrm{R}}$) is the lowest-energy state in the R sector with energy $E_0^{\mathrm{R}} + m$ ($m$ is the single-particle energy of $\eta^{\dag}_0$). Thus, the universal gap function between two sectors is given by
\begin{align}
\Delta(t) &\equiv\frac{L}{2\pi}[(E_0^{\mathrm{R}}+m)-E_0^{\mathrm{NS}}] \nonumber \\
&=\frac{1}{2}[\gamma_{+}(t)-\gamma_{-}(t)]+t \nonumber \\
&=\frac{1}{8}+\frac{t}{2}+t^2\ln 2 +\sum_{l=2}^\infty t^{2l} C_{
\frac{1}{2}}^l (2^{2l-1}-2)\zeta(2l-1) \, ,
\end{align}
which is indeed independent of the short-distance cutoff $a$. This result agrees with Refs.~\cite{Henkel1987,Saleur1987,Oshikawa2019}.

\subsection{Torus partition function}

By using the diagonalized Hamiltonian in both sectors and taking into account that the NS (R) sector has an even (odd) number of fermions, the torus partition function for the Hamiltonian \eqref{eq:Major-QFT} is calculated as follows:
\begin{align}
Z^{\mathcal{T}}(L,\beta,t) &= \mathrm{tr}(e^{-\beta H}) \nonumber \\
&= \mathrm{tr}_{\mathrm{NS}}(e^{-\beta H_{\mathrm{NS}}}) + \mathrm{tr}_{\mathrm{R}}(e^{-\beta H_{\mathrm{R}}}) \nonumber \\
&= \mathrm{tr}\left[ \frac{1+(-1)^{\sum_{k \in \mathrm{NS}}\eta^{\dag}_{k}\eta_{k}}}{2} e^{-\beta \sum_{k\in\mathrm{NS}}\epsilon_k(\eta^{\dag}_{k}\eta_{k} - 1/2)} \right]
+ \mathrm{tr} \left[ \frac{1-(-1)^{\sum_{k \in \mathrm{R}}\eta^{\dag}_{k}\eta_{k}}}{2} e^{-\beta \sum_{k\in\mathrm{R}}\epsilon_k(\eta^{\dag}_{k}\eta_{k} - 1/2)} \right] \nonumber \\
&=\frac{1}{2}[D_{--}(L,\beta,t)+ D_{-+}(L,\beta,t)+ D_{+-}(L,\beta,t)-D_{++}(L,\beta,t)]
\label{eq:Zt}
\end{align}
with
\begin{align}
D_{--}(L,\beta,t)
&= \mathrm{tr}\left[ e^{-\beta \sum_{k\in\mathrm{NS}}\epsilon_k(\eta^{\dag}_{k}\eta_{k} - 1/2)} \right]
= e^{-\pi \frac{\beta}{L}\gamma_{-}(t)} \left|\prod_{n=1}^{\infty}(1+q^{\sqrt{(n-\frac{1}{2})^2+t^2}})\right|^2 \, , \\
D_{-+}(L,\beta,t)
&= \mathrm{tr}\left[ (-1)^{\sum_{k \in \mathrm{NS}}\eta^{\dag}_{k}\eta_{k}} e^{-\beta \sum_{k\in\mathrm{NS}}\epsilon_k(\eta^{\dag}_{k}\eta_{k} - 1/2)} \right]
= e^{-\pi \frac{\beta}{L}\gamma_{-}(t)}\left|\prod_{n=1}^{\infty}(1-q^{\sqrt{(n-\frac{1}{2})^2+t^2}})\right|^2 \, , \\
D_{+-}(L,\beta,t)
&= \mathrm{tr}\left[ e^{-\beta \sum_{k\in\mathrm{R}}\epsilon_k(\eta^{\dag}_{k}\eta_{k} - 1/2)} \right]
= e^{-\pi \frac{\beta}{L}\gamma_{+}(t)}(1+q^t)\left|\prod_{n=1}^{\infty}(1+q^{\sqrt{n^2+t^2}})\right|^2 \, , \\
D_{++}(L,\beta,t)
&= \mathrm{tr}\left[ (-1)^{\sum_{k \in \mathrm{R}}\eta^{\dag}_{k}\eta_{k}} e^{-\beta \sum_{k\in\mathrm{R}}\epsilon_k(\eta^{\dag}_{k}\eta_{k} - 1/2)} \right]
= e^{-\pi \frac{\beta}{L}\gamma_{+}(t)}(1-q^t)\left|\prod_{n=1}^{\infty}(1-q^{\sqrt{n^2+t^2}})\right|^2 \, ,
\end{align}
and $q=e^{-2\pi \frac{\beta}{L}}$. This result agrees with the path integral derivation in Ref.~\cite{Saleur1987}.

The decomposition of the torus partition function $Z^{\mathcal{T}}(L,\beta,t)$ into four terms reflects the well known ``spin structure'' of the Ising field theory in the Majorana fermion formulation. For $D_{\eta \rho}(L,\beta,t)$, the subscripts take four choices, i.e., $\eta = \pm$ and $\rho = \pm$, where $\eta$ ($\rho$) corresponds to the spatial (imaginary-time) direction, and $-$ ($+$) indicates antiperiodic (periodic) boundary condition of the Majorana fermion.

The torus partition function is invariant under modular transformations, regardless of whether it describes the massless critical point or off-critical phases~\cite{Saleur1987}. For our purpose, it is sufficient to consider the modular $\mathcal{S}$ transformation (i.e., the space-time rotation) generated by $L \Longleftrightarrow \beta$, under which $Z^\mathcal{T}$ is unchanged, i.e., $Z^{\mathcal{T}}(L,\beta,m)=Z^{\mathcal{T}}(\beta,L,m)$ ($m$: Majorana mass). This can be easily understood with a classical Ising model with isotropic couplings on a square lattice, with length $L\sim N_x$, width $\beta\sim N_y$, and mass $m\sim T-T_c$. The torus partition function is obviously invariant under exchange of $N_x$ and $N_y$.

Under the space-time rotation, the dimensionless mass transforms as $t = \frac{m L}{2\pi} \Longleftrightarrow s = \frac{m \beta}{2\pi}$. Together with the spin structure described above, we obtain the following useful modular transformation property for $D_{\eta \rho}(L,\beta,t)$:
\begin{align}
D_{\eta \rho}(L,\beta,t) = D_{\rho \eta }(\beta,L,s) \, .
\label{eq:massive-modular}
\end{align}

\subsection{Klein bottle entropy as a universal scaling function}

The Klein bottle partition function is defined by $Z^{\mathcal{K}}=\mathrm{tr}(\Omega e^{-\beta H})$. As the Hamiltonian \eqref{eq:Major-QFT} is already diagonalized, it is most convenient to consider the action of $\Omega$ on the energy eigenbasis. For the fermionic basis of the Ising theory, the states that are invariant under the action of $\Omega$ are as follows~\cite{Tu2017,Tang2019}: (i) The ground states in the NS and R sectors, i.e., $\Omega |0\rangle_{\mathrm{NS}} = |0\rangle_{\mathrm{NS}}$ and $\Omega \eta^{\dag}_0 |0\rangle_{\mathrm{R}} = \eta^{\dag}_0|0\rangle_{\mathrm{R}}$. (ii) Higher-energy states that are ``left-right symmetric'', such as the two-fermion state $\eta^{\dag}_k \eta^{\dag}_{-k} |0\rangle_{\mathrm{NS}}$ in the NS sector and the three-fermion state $\eta^{\dag}_k \eta^{\dag}_{-k} \eta^{\dag}_0 |0\rangle_{\mathrm{R}}$ ($k \neq 0$). Other states that are not ``left-right symmetric'' under $\Omega$, such as $\eta^{\dag}_k \eta^{\dag}_{-k'} |0\rangle_{\mathrm{NS}}$ with $k \neq k'$ ($\Omega \eta^{\dag}_k \eta^{\dag}_{-k'} |0\rangle_{\mathrm{NS}} \propto \eta^{\dag}_{k'} \eta^{\dag}_{-k} |0\rangle_{\mathrm{NS}}$), would not contribute to the Klein bottle partition function.

Taking these into account, the Klein bottle partition function is evaluated as
\begin{align}
Z^{\mathcal{K}}(L,\beta,t) &=\mathrm{tr}(\Omega e^{-\beta\hat{H}}) \nonumber \\
&= \mathrm{tr}_{\mathrm{NS}}(\Omega e^{-\beta H_{\mathrm{NS}}}) + \mathrm{tr}_{\mathrm{R}}(\Omega e^{-\beta H_{\mathrm{R}}}) \nonumber \\
&= \mathrm{tr}_{\mathrm{NS}}\left[\Omega  e^{-\beta \sum_{k\in\mathrm{NS}}\epsilon_k(\eta^{\dag}_{k}\eta_{k} - 1/2)} \right] + \mathrm{tr}_{\mathrm{R}}\left[\Omega e^{-\beta \sum_{k\in\mathrm{R}}\epsilon_k(\eta^{\dag}_{k}\eta_{k} - 1/2)} \right] \nonumber \\
&= D_{\mathrm{NS}}(L,\beta,t)+D_{\mathrm{R}}(L,\beta,t)
\label{eq:Zk}
\end{align}
with
\begin{align}
D_{\mathrm{NS}}(L,\beta,t) &= e^{-\pi \frac{\beta}{L}\gamma_{-}(t)} \prod_{n=1}^\infty(1+q^{2\sqrt{{(n-\frac{1}{2})}^2+t^2}})=\sqrt{D_{--}(L,2\beta,t)} \, , \nonumber \\
D_{\mathrm{R}}(L,\beta,t) &= e^{-\pi \frac{\beta}{L}\gamma_{+}(t)} q^t\prod_{n=1}^\infty(1+q^{2\sqrt{n^2+t^2}})=\frac{q^t}{\sqrt{1+q^{2t}}}\sqrt{D_{+-}(L,2\beta,t)} \, ,
\end{align}
where $q=e^{-2\pi \frac{\beta}{L}}$.

To determine the Klein bottle entropy in the limit $L \gg \beta$, we perform the modular $\mathcal{S}$ transformation ($L \Longleftrightarrow \beta$), so that $t=\frac{mL}{2\pi} \Longrightarrow s=\frac{m\beta}{2\pi}$ and $q=e^{-2\pi \frac{\beta}{L}} \rightarrow 1 \Longrightarrow q'=e^{-2\pi \frac{L}{\beta}} \rightarrow 0$. By using the transformation property of $D_{\eta\rho}$ [see Eq.~\eqref{eq:massive-modular}], the Klein bottle partition function $Z^{\mathcal{K}}(2L,\frac{\beta}{2},m)$ (for the partition function with length $2L$ and mass $m$, the dimensionless mass is $\frac{2L\cdot m}{2\pi}=2t$ by definition) can be calculated as
\begin{align}
Z^{\mathcal{K}}(2L,\frac{\beta}{2},m) &=Z^{\mathcal{K}}(2L,\frac{\beta}{2},2t) \nonumber \\
&=D_{\mathrm{NS}}(2L,\frac{\beta}{2},2t)+D_{\mathrm{R}}(2L,\frac{\beta}{2},2t) \nonumber \\
&=\sqrt{D_{--}(2L,\beta,2t)}+\frac{q^{\frac{t}{2}}}{\sqrt{1+q^t}}\sqrt{D_{+-}(2L,\beta,2t)} \nonumber \\
&=\sqrt{D_{--}(\beta,2L,s)}+\frac{e^{-\pi s}}{\sqrt{1+e^{-2\pi s}}}\sqrt{D_{-+}(\beta,2L,s)} \, ,
\end{align}
and the torus partition function as
\begin{align}
Z^\mathcal{T}(L,\beta,t)=Z^\mathcal{T}(\beta,L,s)=\frac{1}{2}[D_{--}(\beta,L,s)+D_{+-}(\beta,L,s)+D_{-+}(\beta,L,s)-D_{++}(\beta,L,s)] \, .
\end{align}

For $L \gg \beta$, we have,
\begin{align}
D_{--}(\beta,2L,s)=e^{-\pi\frac{2L}{\beta}\gamma_{-}(s)} \, , \nonumber \\
D_{-+}(\beta,2L,s)=e^{-\pi\frac{2L}{\beta}\gamma_{-}(s)} \, , \nonumber \\
Z^\mathcal{T}(\beta,L,s) = e^{-\pi\frac{L}{\beta}\gamma_{-}(s)} \, .
\end{align}
Since $\gamma_-(s)<\gamma_+(s)$, when $L\gg\beta$, the contributions of $D_{+-}(\beta,L,s)$ and $D_{++}(\beta,L,s)$ to $Z^\mathcal{T}(\beta,L,s)$ are exponentially suppressed.

Thus, for the Ising CFT with thermal perturbation, the Klein bottle entropy as a universal scaling function is given by
\begin{equation}
K(s)=\lim_{L\rightarrow \infty}\frac{Z^{\mathcal{K}}(2L,\frac{\beta}{2},m)}{Z^{\mathcal{T}}(L,\beta,m)}=1+\frac{1}{\sqrt{1+e^{2\pi s}}} \, .
\label{eq:Ks-mass}
\end{equation}

\end{document}